\documentclass{cflowproc}

\usepackage{natbib}

\begin{document}


\shortauthors{Blanton}     
\shorttitle{Radio Source / X-ray Gas Interactions} 

\title{The Interaction of Radio Sources and X-ray-Emitting Gas in Cooling
Flows}   

\author{Elizabeth L. Blanton\affilmark{1,2}}   

\affil{1}{University of Virginia, Department of Astronomy, P.O. Box 3818,
Charlottesville, VA 22903; eblanton@virginia.edu}   
\and                                
\affil{2}{{\it Chandra} Fellow}


\begin{abstract}
Recent observations of the interactions between radio sources and the 
X-ray-emitting gas in cooling flows in the cores of clusters of galaxies are
reviewed. The radio sources inflate bubbles in the X-ray gas, which then rise 
buoyantly outward in the clusters transporting energy to the intracluster 
medium 
(ICM). The bright rims of gas around the radio bubbles are cool, rather than 
hot, and do not show signs of being strongly shocked. Energy deposited into 
the ICM 
over the lifetime of a cluster through several outbursts of a radio source
helps to account for at least some of the gas that is missing in cooling 
flows at low temperatures. 
\end{abstract}


\section{Introduction}
\label{blanton:intro}

The vast majority of cooling flow clusters contain powerful radio sources
associated with central cD galaxies.
Initial evidence of radio sources displacing, and evacuating cavities in, 
the X-ray-emitting intracluster medium (ICM) was found with {\it ROSAT} 
observations of a few sources including Perseus (B\"ohringer et al.\ 1993), 
Abell 4059 (Huang \& Sarazin 1998), and Abell 2052 (Rizza et al.\ 2000).  
Models predicted that the radio sources would shock the ICM, and that
the X-ray emission surrounding the lobes would appear hot (Heinz, Reynolds,
\& Begelman 1998).
High-resolution images from {\it Chandra} have revealed many more cases
of radio sources profoundly effecting the ICM
by displacing it and creating X-ray deficient ``holes'' or ``bubbles.''  
The {\it Chandra} data
allow us to study the physics of the interaction in much more 
detail (i.e. Hydra A, McNamara et al.\ 2000; Perseus, Fabian et al.\ 2000;
Abell 2052, Blanton et al.\ 2001; Abell 2597, McNamara et al.\ 2001;  
Abell 496,
Dupke \& White, 2001; MKW 3s, Mazzotta et al.\ 2002; RBS797, Schindler et al.\
2001; Abell 2199, Johnstone et al.\ 2002; Abell 4059, Heinz et al.\ 2002;
Virgo, Young et al.\ 2002;
Centaurus, Sanders \& Fabian 2002; Cygnus A, Smith et al.\ 2002; Abell 478,
Sun et al.\ 2003).

A long-standing problem with cooling flow models has been that the mass of
gas measured to be cooling from X-ray temperatures, based on 
surface-brightness and spectral
studies with {\it Einstein} and {\it ROSAT}
has not been detected in sufficient quantities at cooler temperatures.
High-resolution spectroscopy with {\it XMM-Newton} provided
direct evidence that gas was cooling in these clusters, but very large 
masses of gas (hundreds of solar masses) were seemingly 
cooling over only a limited range of temperatures.  Emission lines such as Fe 
XVII expected from gas cooling below approximately 2 keV were not detected
and in general, most of the ICM seems to cool to about one-half to 
one-third of the cluster ambient temperature (Peterson et al.\ 2003).
Several possible solutions have been proposed for the lack of cool 
X-ray gas seen in the new observations (Fabian et al.\ 2001;
Peterson et al.\ 2001).  These include mixing, heating by central 
active-galactic nuclei (AGN), 
inhomogeneous abundances, and differential absorption.
Heating of the gas by a central radio source has also been discussed 
recently by
B\"ohringer et al.\ (2002), Churazov et al.\ (2002), Ruszkowski \& Begelman
(2002), Kaiser \& Binney (2003), and Br\"uggen (2003), among others. 

A common finding with the {\it Chandra} data is 
X-ray deficient holes that correspond with radio emission from the lobes of
a central AGN. These
holes are typically surrounded by bright shells of dense, X-ray-emitting gas.
One of the surprises of the {\it Chandra} observations is that the 
X-ray-bright rims surrounding the radio sources observed in cooling flow 
clusters
were found to be cooler, rather than hotter, than the neighboring cluster
gas (i.e. Perseus, Schmidt et al.\ 2002; Hydra A, Nulsen et al.\ 2002;
Abell 2052, Blanton et al.\ 2003).  
The bright shells show no evidence of current strong shocks.
Soker, Blanton, \& Sarazin (2002) found that for the case of Abell 2052,
the morphology was well-explained by weak shocks occurring in the past, and 
strong shocks were generally ruled out.
To date, no case of current strong-shock heating of the ICM in a cooling flow
cluster by an AGN has been observed.  Although strong-shock 
heating from the radio source as proposed by Heinz, Reynolds, \& Begelman
(1998) and Rizza et al.\ (2000) may not be the explanation
for the fate of the missing cool gas, energy input
from a radio source in the form of weak shocks 
(e.g. Reynolds, Heinz, \& Begelman 2002) and buoyantly rising bubbles of
relativistic plasma (e.g. Churazov et al.\ 2002) can still contribute to 
heating.  

\section{{\it Chandra} Observations of Bubbles and Temperature Structure}
\label{blanton:bubb_temp}

\begin{figure}
\plotone{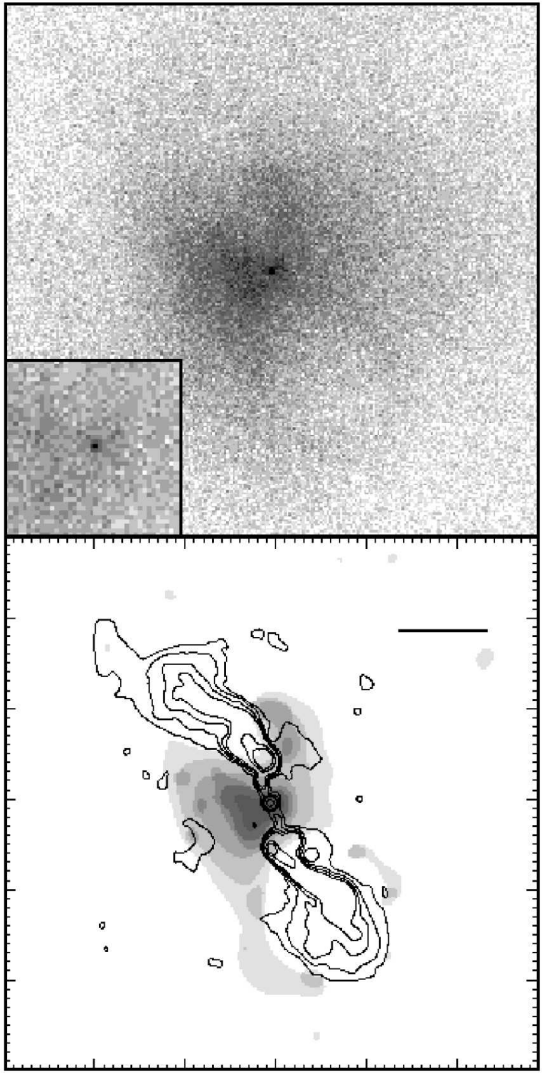}
\figcaption{The first radio source / cluster cooling flow interaction
observed with {\it Chandra}:  Hydra A (McNamara et al.\ 2000).
\label{blanton:hydra1}}
\end{figure}

\begin{figure}
\plotone{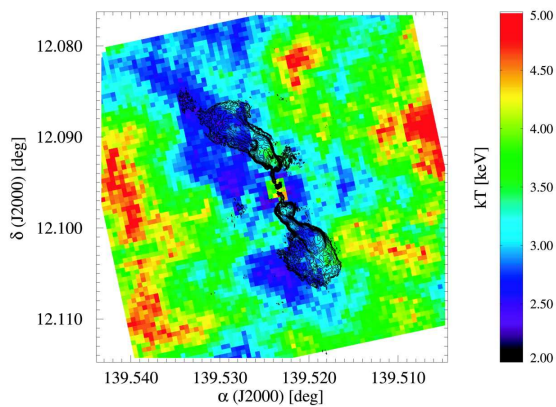}
\figcaption{The temperature map of the Hydra A cluster, with radio
contours superposed (Nulsen et al.\ 2002).  The coolest gas is
found in the regions surrounding the radio source, and there is no
evidence for current strong-shock heating of the ICM from the radio
lobes.
\label{blanton:hydra2}}
\end{figure}

\begin{figure}
\plotone{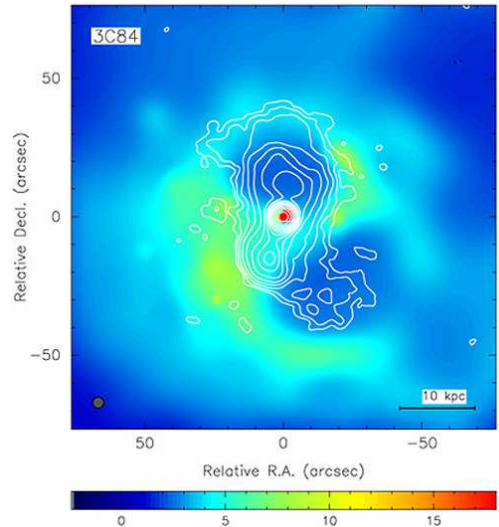}
\figcaption{Overlay of radio contours onto the adaptively-smoothed 
{\it Chandra} X-ray image of the Perseus cluster.  Radio:NSF/AURA/VLA; 
X-ray:NASA/IoA/A.Fabian et al.
\label{blanton:perseus1}}
\end{figure}

\begin{figure}
\plotone{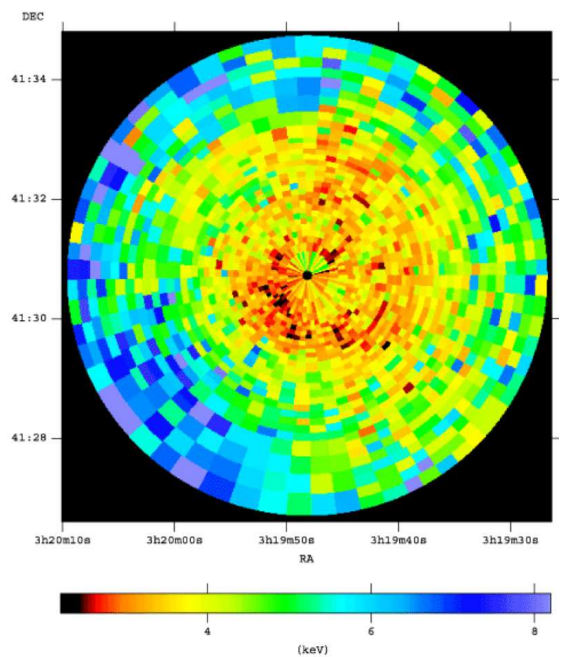}
\figcaption{Temperature map of the central regions of the Perseus cluster
(Schmidt et al.\ 2002).
\label{blanton:perseustmap}}
\end{figure}

The first cooling flow cluster with a central radio source observed 
by {\it Chandra} was Hydra A (McNamara et al.\ 2000; David et al.\ 2001,
Nulsen et al.\ 2002).  Hydra A is a 
powerful, doubled-lobed, FR I radio source (3C 218), and the radio lobes were
found to be anti-correlated with the cluster gas.  The cavities evacuated
by the radio source are approximately 25 kpc in diameter.  An image
of the radio source / X-ray gas interaction is shown in Figure 
\ref{blanton:hydra1}, and is from McNamara et al.\ (2000).
The cooling time in the center of the cluster is $6 \times 10^8$ yr, and
surprisingly, the {\it coolest} gas was found in the regions surrounding 
the lobes.  This is seen in the temperature map from Nulsen et al.\ (2002),
shown here in Figure \ref{blanton:hydra2}.  There is currently no evidence
that the radio source is strongly shocking the ICM, but weak shocks are 
not ruled out with a limit on the Mach number of ${\cal M} < 1.23$.
The cooling of the ICM is found to occur only over a limited temperature
range, and repeated outbursts from the radio source would be necessary to
prevent cooling to lower temperatures (David et al.\ 2001).

The Perseus cluster (Abell 426) was first observed with {\it Chandra} in 
early 2000 (Fabian et al.\ 2000).  This nearby cooling flow cluster 
($z = 0.0183$) is the brightest cluster in the X-ray sky and
contains the powerful double-lobed radio source 3C 84 in the central
galaxy, NGC 1275.  Perseus provides one of the clearest examples of the
interaction between the central radio source and the X-ray-emitting ICM,
and signs of the interaction were already seen in the {\it ROSAT} data
(B\"ohringer et al.\ 1993).
The cluster exhibits two distinct bubbles in the X-ray gas that are filled 
with 1.4 GHz radio plasma and surrounded by bright shells of X-ray emission.
The adaptively-smoothed {\it Chandra} image of the center of Perseus, with 
radio contours superposed, is shown in Fig.\ \ref{blanton:perseus1} (Fabian
et al.\ 2000).
The cooling time in the cluster center is approximately $10^8$ yr, and no
evidence for current strong-shock heating is seen in the temperature map
(Fig.\ \ref{blanton:perseustmap}, Schmidt et al.\ 2002).  The regions 
surrounding the radio source are the coolest in the X-ray. 
Presented at this meeting were initial exciting results from a much deeper 
observation of the Perseus cluster that was performed by {\it Chandra}
in 2003, revealing intriguing ripple features in the X-ray surface brightness
that are interpreted as resulting from the propagation of weak shocks
and viscously-dissipating sound waves into the ICM and resulting from
repeated outbursts of the central radio source (Fabian, this conference;
Fabian et al.\ 2003).

\begin{figure}
\plotone{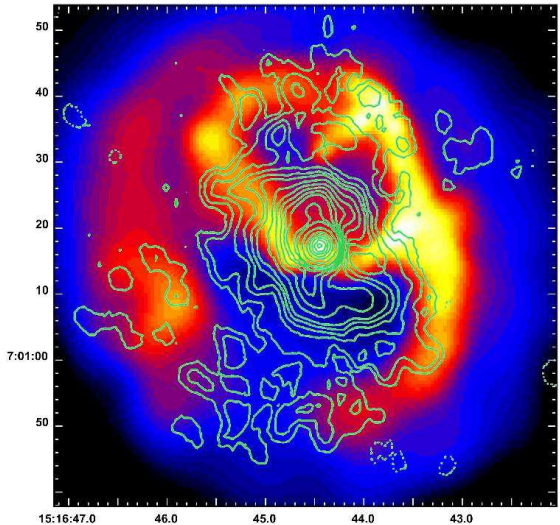}
\figcaption{Overlay of 1.4 GHz radio contours onto the adaptively-smoothed
{\it Chandra} image of Abell 2052 (Blanton et al.\ 2001, 2003).
\label{blanton:a2052rad}}
\end{figure}

\begin{figure}
\plotone{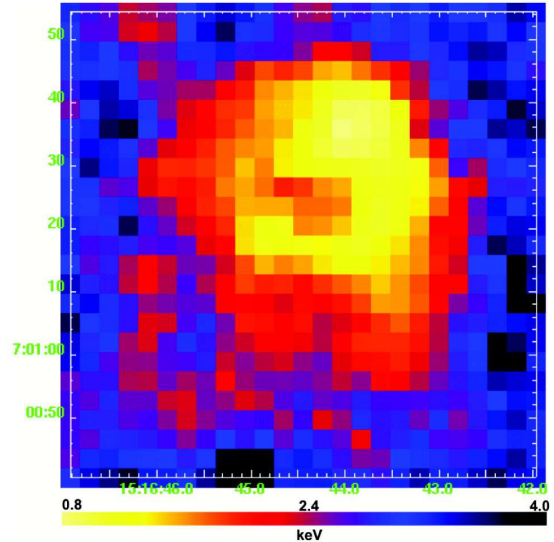}
\figcaption{Temperature map of the central region of Abell 2052 derived from
the {\it Chandra} data (Blanton et al.\ 2003).  The shells surrounding the
radio source are cool and show no signs of strong-shock heating.
\label{blanton:a2052tmap}}
\end{figure}

\begin{figure}
\plotone{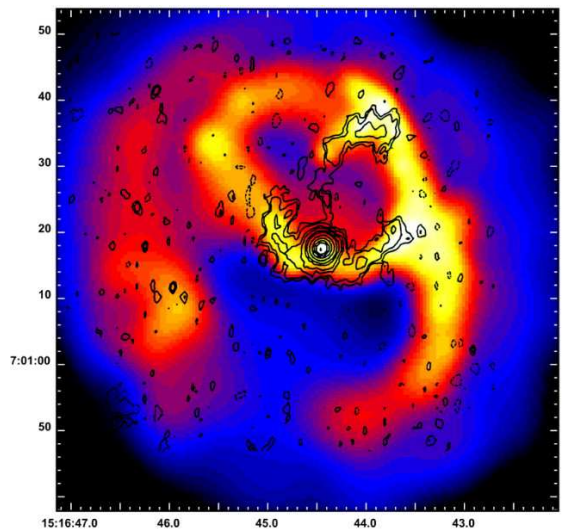}
\figcaption{Overlay of optical emission line contours from Baum et al.\ 
(1988) onto the adaptively-smoothed {\it Chandra} image of the center of
Abell 2052 (Blanton et al.\ 2001).
\label{blanton:a2052ha}}
\end{figure}

Abell 2052 shows structures in its center that are very similar to those
in Perseus.  This cluster is nearby ($z=0.0348$), and the central cD galaxy
is host to the powerful, double-lobed FR I radio source 3C 317.  The 
{\it Chandra} image (Blanton et al.\ 2001, 2003) revealed clear deficits in 
the X-ray emission to the north and south of the cluster center that are
filled with 1.4 GHz radio emission (Fig.\ \ref{blanton:a2052rad}).  The 
bubbles are approximately 20 kpc in diameter and are
surrounded by bright shells of X-ray emission.  Extrapolating
the smooth density profile of the cluster outside of the shells into the
center of the cluster shows that the mass of gas that would have filled the 
holes is consistent with the mass of gas currently found in the shells. This
confirms the visual impression that the ICM has been swept aside by the 
radio lobes and compressed into the X-ray-bright shells.  The temperature map
(Fig.\ \ref{blanton:a2052tmap}) reveals that the shells surrounding the radio 
lobes are cool, and show
no evidence of being strongly shocked, with a limit to the Mach number of
${\cal M} < 1.2$.  The cooling time in the shells is $2.6 \times 10^8$ yr.
An overlay of optical emission line contours of H$\alpha$+[\ion{N}{2}]
onto the X-ray image (Fig.\ \ref{blanton:a2052ha}) reveals a striking 
positive correlation between the brightest parts of the X-ray shells and the 
optical line emission.  The optical emission represents gas with a temperature
of $\approx 10^4$ K, and the temperature of the X-ray gas in these same
regions is $\approx 10^7$ K, so at least some gas is cooling to low 
temperatures in this cooling flow cluster.  
The cooling time of the shells is approximately an order of magnitude longer
than the radio source lifetime of $10^7$ yr, and therefore, the majority of
the cooling to
low temperatures in the X-ray shells likely occurred when the gas was closer
to the cluster center and was subsequently pushed outward by the radio source.

\begin{figure}
\plotone{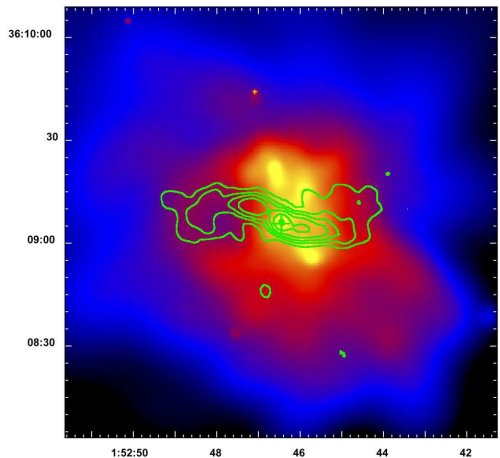}
\figcaption{Overlay of 1.4 GHz radio contours (Parma et al.\ 1986) onto 
the adaptively-smoothed {\it Chandra} image of Abell 262 (Blanton et al.,
in preparation).
\label{blanton:a262rad}}
\end{figure}

\begin{figure}
\plotone{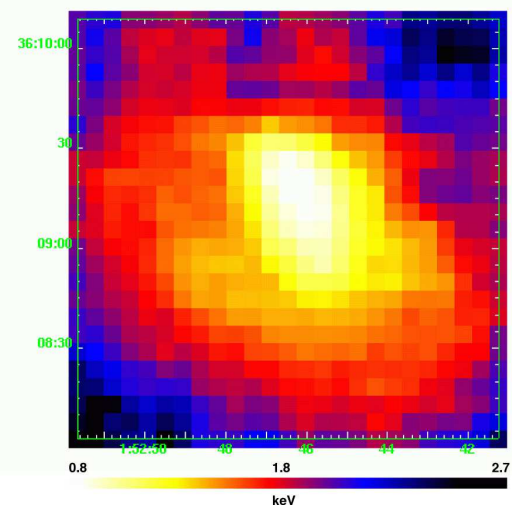}
\figcaption{Temperature map of the central region of Abell 262
(Blanton et al., in preparation).
\label{blanton:a262tmap}}
\end{figure}

Another example of radio source / X-ray gas interaction in a cooling
flow cluster observed with {\it Chandra} is Abell 262 (Blanton et al., in
preparation).  This cluster is
at a redshift of $z=0.0163$ in the same supercluster as Perseus, but is
less luminous than Perseus in the X-ray, and has a smaller cooling flow.
The double-lobed radio source is orders of magnitude less powerful than those
in the previous examples, with a power at 1.4 GHz of only 
$P_{1.4} = 4.7 \times 10^{22}$ W Hz$^{-1}$ (Parma et al.\ 1986).
Still, the radio source has blown a bubble in the ICM, clearly seen to the
east of the cluster center (Fig.\ \ref{blanton:a262rad}).  This bubble is much
smaller than those in the other clusters discussed above, with a diameter
of 5 kpc compared to $20 - 25$ kpc for the others.
Similar to the other cases, the cooling time for the gas surrounding the 
radio lobes is $3 \times 10^8$ yr, and the temperature map shows no evidence
of strong-shock heating (Fig.\ \ref{blanton:a262tmap}).

\section{Evidence of Shock Heating in Galaxies}
\label{blanton:shockgal}

As described in the previous section, there is currently no direct evidence
for strong-shock heating in cooling flow clusters from the current outburst 
of the
radio source.  The temperature maps reveal that the compressed ICM in the 
shells surrounding the radio bubbles is cool.

\begin{figure}
\plotone{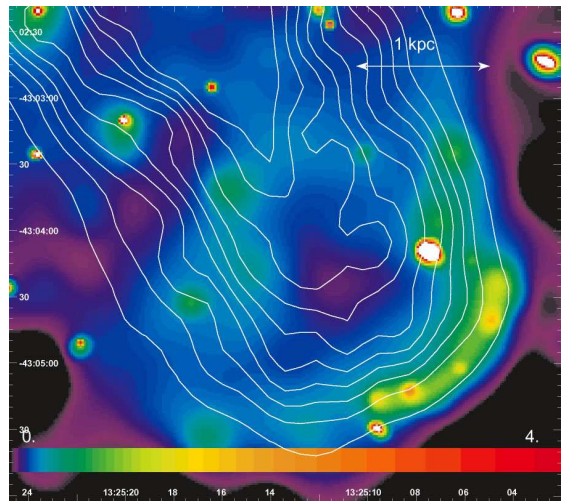}
\figcaption{Adaptively-smoothed {\it Chandra} image of the SW lobe of Cen A, 
with radio contours superposed (Kraft et al.\ 2003).  The bright cap of
emission is thought to be ISM that has been shocked by the radio source.
\label{blanton:cena}}
\end{figure}

It is worth mentioning here examples of what appears to be shock 
heating of X-ray-emitting gas by a radio source.  The first, and
more obvious case, is that of the nearby radio galaxy, Centaurus A.  
This E galaxy was observed with {\it Chandra} and {\it XMM-Newton}, as 
described in Kraft et al.\ (2003).  It is the nearest active galaxy, at a 
distance of 3.4 Mpc.  The radio source is a double-lobed FR I with a power
of $1.9 \times 10^{24}$ W Hz$^{-1}$ at 1.4 GHz.  There is a shell or 
cap of emission
seen in the X-ray along the edge of the SW radio lobe.  This feature was
interpreted as shocked interstellar medium (ISM), and exhibits a 
temperature and pressure that are higher than its surroundings.  The pressure
jump is consistent with a shock with a Mach number of ${\cal M} = 8.5$.
The brightened cap of emission is shown in Fig.\ \ref{blanton:cena} (Kraft
et al.\ 2003).  While the shocked gas here is ISM, rather than ICM, it still
represents the type of signature we might expect to see in cluster gas that
has been shocked by a radio source.

Another case is the elliptical galaxy NGC 4636.  This galaxy is located in
the outer part of the Virgo cluster.  The {\it Chandra} image (Jones et al.\
2002) shows bright arm-like features with sharp edges.  These arm-like 
features have higher temperatures and pressures than the surrounding gas,
consistent with gas that experienced a shock with Mach number
${\cal M} = 1.73$.  However, there is no strong radio source that currently
corresponds with the arm-like features, and they may or may not result from
a previous radio outburst.

\section{Pressure in X-ray Shells}
\label{blanton:pressure}

A common feature of the X-ray-bright shells surrounding the radio bubbles
shown in \S\ref{blanton:bubb_temp} is that the pressure measured for the
shells is approximately equal to that just outside of them.  In other words,
there is no evidence for a strong shock.  Another feature common to many
of these objects is that
the pressure measured in the X-ray-bright shells is about an order of 
magnitude higher than the pressure measured from the radio data within the 
radio-bright bubbles,
assuming equipartition of energies (an example is Abell 2052, with an X-ray
shell pressure of $1.5 \times 10^{-10}$ dyn cm$^{-2}$ [Blanton et al.\ 2001],
and a radio
equipartition pressure of $2 \times 10^{-11}$ dyn cm$^{-2}$ 
[Zhao et al.\ 1993]).  However, we expect that the bubbles
and the shells are in pressure equilibrium, since otherwise they would
collapse and fill in.  Therefore, either some of the assumptions
made for the equipartition pressure estimates are incorrect, or there is 
an additional source of pressure within the radio bubbles.  This additional
pressure component may be magnetic fields, low energy relativistic electrons,
or very hot, diffuse, thermal gas that would not be detected by {\it Chandra}
because of its low surface brightness in the {\it Chandra} energy band.
The temperature of hot, thermal gas that would provide the required
pressure to support the X-ray shells has been limited to $> 15$ keV for 
Hydra A (Nulsen et al.\ 2002), $> 11$ keV for Perseus (Schmidt et al.\ 2002), 
and $> 20$ keV for Abell 2052 (Blanton et al.\ 2003).  High sensitivity at 
high energies is necessary to detect diffuse gas at such temperatures, and 
{\it XMM-Newton} or the upcoming {\it Constellation-X} may be able to 
detect it.

A detection of gas within an X-ray depression with a temperature significantly
hotter than its surroundings has been made using {\it Chandra} data of  
the cooling flow cluster
MKW 3s (Mazzotta et al.\ 2002).  The gas in the bubble is hotter than the
gas at any radius in the cluster, and the temperature measurement is therefore
not a projection effect.  The deprojected gas temperature within the bubble
is 7.5 keV, compared with a temperature of $3.5 - 4$ keV for the surrounding
emission.  This cluster contains a central radio source, however the 1.4
GHz radio emission is not directly connected with the X-ray depression,
as shown in Mazzotta et al.\ 2002.

\section{Buoyantly Rising Bubbles}
\label{blanton:buoyant}

\begin{figure}
\plotone{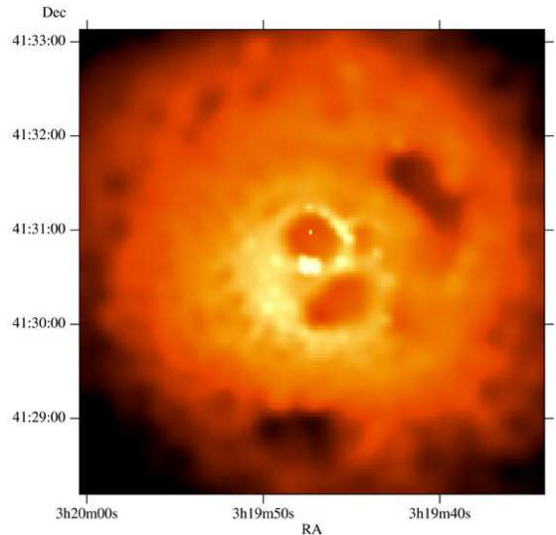}
\figcaption{Adaptively-smoothed {\it Chandra} image of the center of the
Perseus cluster.  The inner bubbles that are associated with the current
1.4 GHz radio emission are seen, as well as two ghost cavities to the S and
NW of the cluster center.  NASA/IoA/A.Fabian et al.
\label{blanton:perseusghost}}
\end{figure}

\begin{figure}
\plotone{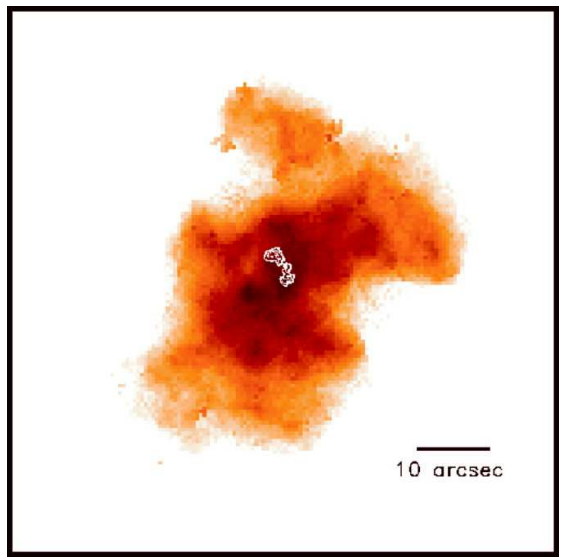}
\figcaption{The central region of Abell 2597 as observed by {\it Chandra}, 
after subtracting a smooth background cluster model.  The contours of the
small central radio source are superposed.  Outer ghost cavities are to the
NE and SW of the cluster center and are not clearly associated with the 
8.44 GHz radio contours shown here.  This figure is from McNamara et al.\
(2001).
\label{blanton:a2597}}
\end{figure}

The density inside the radio bubbles is much lower than that of the ambient
gas, so the bubbles should be buoyant and rise outward in the clusters.
These rising bubbles can transport energy and magnetic fields into the 
clusters.

Evidence of bubbles that have risen buoyantly away from the cluster centers
has been found in the Perseus (Fabian et al.\ 2000) and Abell 2597 (McNamara
et al.\ 2001) clusters.  These bubbles are not spatially coincident with
the 1.4 GHz (Perseus) or 8.44 GHz (Abell 2597) radio contours from the 
central AGN, and are farther from
the cluster center than the radio emission.  These features have
been referred to as ``ghost bubbles'' or ``ghost cavities'' and are thought
to result from a previous outburst of the radio source.  Examples of these
features are shown in Figures \ref{blanton:perseusghost} (Perseus,
Fabian et al.\ 2000) and \ref{blanton:a2597} (Abell 2597, McNamara et al.\
2001).
The buoyancy rise time for the ghost cavities to arrive at their projected
positions has been calculated for these objects, and this timescale reveals
the repetition rate of the radio outbursts from the AGN, assuming that 
the cavities result from a previous outburst.  For A2597, for example,
the repetition rate is approximately $10^8$ yr.  This is similar to
the cooling time of the central gas, suggesting that a feedback process
is operating, where cooling gas fuels the AGN, the AGN has an outburst and
heats the gas, then the gas cools and fuels the AGN, etc. (McNamara et al.\
2001).

Further evidence that the ghost cavities were created by radio lobes earlier 
in the life of the AGN comes from the detection of low frequency radio
emission that is spatially coincident with the outer cavities.  The clearest
example of this correlation is found in the Perseus cluster when the
74 MHz radio emission is compared with the X-ray emission (Fabian et al.\
2002).  There also appears to be a similar correlation in the Abell 2597
cluster, based on the comparison of lower frequency radio data than that shown
in Fig.\ \ref{blanton:a2597} and the {\it Chandra} image 
(McNamara et al.\ 2001).

\begin{figure}
\plotone{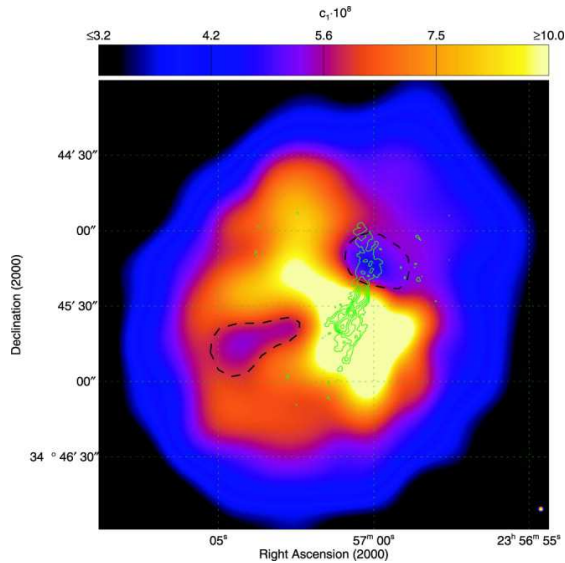}
\figcaption{Adaptively-smoothed {\it Chandra} image of the cooling flow
cluster Abell 4059.  Contours of the 8 GHz radio emission are superposed.
The radio emission only partly fills the cavities in the X-ray (Heinz et al.\
2002).
\label{blanton:a4059}}
\end{figure}

\section{Intermediate Cases}
\label{blanton:intermed}

\begin{figure}
\plotone{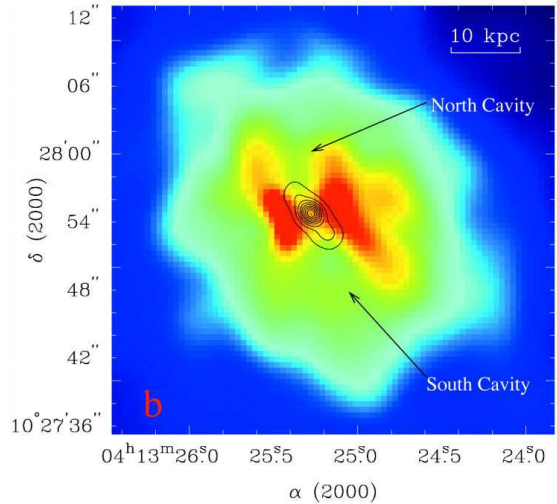}
\figcaption{Adaptively-smoothed {\it Chandra} image of the cooling flow
cluster Abell 478, with contours of the 1.4 GHz radio emission overlaid.
This is a similar case to that of Abell 4059, where the radio emission
only partly fills the X-ray holes (Sun et al.\ 2003).
\label{blanton:a478}}
\end{figure}

In the example cooling flow clusters with central radio sources described 
above, there is either a clear correlation between the radio data and
the X-ray emission, such that the radio emission fills the bubbles in the
X-ray, or the bubbles are completely devoid of (high frequency) radio
emission (the ghost cavities).
In addition to these examples, there also exist intermediate cases, 
``missing links'' as suggested by Heinz et al.\ 2002, where the radio 
emission partly fills the holes seen
in the X-ray.  In these cases, it is possible that the radio emission that
previously filled the cavities has faded, due to synchrotron losses of the
relativistic electrons.
Good examples of these intermediate cases are Abell 4059 (Fig.\ 
\ref{blanton:a4059}, Heinz et al.\ 2002) and Abell 478 (Fig.\ 
\ref{blanton:a478}, Sun et al.\ 2003).

\bigskip
\bigskip
\bigskip
\section{Entrainment of Cool Gas}
\label{blanton:cool}

Buoyantly rising bubbles transport energy and magnetic fields outward into
clusters.  In addition, these rising bubbles create channels in the ICM, and
entrain cool cluster gas from the center outward, where it will be mixed
with hotter gas.

This type of entrainment is seen clearly in M87 in the Virgo cluster
(Young et al.\ 2002).  In this system, the X-ray temperature map reveals that
an arc of cool gas follows the same path as the radio lobes.
The metallicity in the arc is somewhat higher than
the surroundings, which is consistent with the picture of the gas in the
arc originating closer to the cluster center, where the average metallicity
is higher than in the outer regions of the cluster.

An additional example is Abell 133 (Fujita et al.\ 2002).  This cluster
includes a radio source that is offset from the center of the cluster.  
The radio 
source was originally thought to be a radio relic, likely produced from
a merger shock.  The {\it Chandra} image revealed a filament connecting the
radio source with the cluster center.  The filament is cool, and shows no
evidence of shocks.  Fujita et al.\ surmise that the radio emission is 
probably a detached lobe from the central AGN.  The lobe may have been 
displaced by
the motion of the cD or through buoyancy.  The filament of X-ray emission
is likely to be cool gas that was entrained outward from the center of the
cluster by the radio source.

\section{Can Radio Sources Offset the Cooling in Cooling Flows?}
\label{blanton:offset}

We have seen that radio sources have a profound effect on the X-ray-emitting
ICM, inflating bubbles that rise buoyantly into the clusters.
But, is the energy deposition into the ICM from the radio sources 
sufficient to account for the lack of gas seen at very low temperatures
in cooling flow clusters?  Using the pressures of the X-ray-bright shells
surrounding the bubbles, and the volumes of the bubbles, we can determine the
total energy output of a radio source.  The energy output includes both
the internal energy of the bubble and the work done to expand the bubble.
From Churazov et al.\ (2002), the total energy required for a radio source
to inflate a bubble in the ICM is
\begin{equation}
E_{\rm rad} = \frac{1}{(\gamma - 1)}PV + PdV = \frac{\gamma}{(\gamma - 1)}PV
\, ,
\end{equation}
where $V$ is the volume of the bubble, and $\gamma$ is the mean adiabatic
index of the fluid in the bubble (5/3 for non-relativistic gas or 4/3 for
relativistic gas).
To get the average rate of energy input from a central radio source, we 
divide this energy by the repetition rate of the radio source, based on the 
buoyancy rise time of ghost cavities, and with a value of approximately
$10^8$ yr (see \S \ref{blanton:buoyant}).
We compare this energy input rate with the luminosity of cooling gas derived
from spectral fitting to the X-ray data.  The luminosity of isobaric cooling 
gas is given by
\begin{equation}
L_{\rm{cool}} = \frac{5}{2}\frac{kT}{\mu m_{p}} \dot{M} ,
\end{equation}
where $kT$ is the temperature of the ICM outside of the cooling region,
$\dot{M}$ is the mass-deposition rate, and $\mu$ is the mean mass per particle in
units of the proton mass.

For the case of Hydra A (McNamara et al.\ 2000; David et al.\ 2001; Nulsen
et al.\ 2002), the total power output of the radio source, derived as 
described above from the X-ray pressure and volumes of the radio cavities,
is $2.7 \times 10^{44}$ erg s$^{-1}$.  The cooling luminosity, using
$\dot{M} = 300$ M$_{\odot}$ yr$^{-1}$ and $kT = 3.4$ keV, is
$L_{\rm{cool}} = 3 \times 10^{44}$ erg s$^{-1}$.  Therefore, in this case,
just based on these simple energy arguments,
the radio source is depositing enough energy into the ICM on average 
to offset the
cooling gas.  A similar test performed for Abell 2052 (Blanton et al.\ 2003)
shows that the radio source also has sufficient power 
($3 \times 10^{43}$ erg s$^{-1}$) to offset the cooling gas
($L_{\rm{cool}} = 3 \times 10^{43}$ erg s$^{-1}$ with 
$\dot{M} = 42$ M$_{\odot}$ yr$^{-1}$ and $kT = 3$ keV).
The situation is different for Abell 262 (Blanton et al., in preparation), 
where the radio source power
($3.4 \times 10^{41}$ erg s$^{-1}$)
falls more than an order of magnitude short of that required to offset
the cooling luminosity ($L_{\rm{cool}} = 1.3 \times 10^{43}$ erg s$^{-1}$,
using $kT = 2.65$ keV, and the mass-deposition rate of 
$18.8~M_{\odot}$ yr$^{-1}$).  However, this radio source is much less luminous
than Hydra A or 3C 317 in Abell 2052, and the bubble volume inflated in 
Abell 262 is
much smaller than that in the others.  The bubble diameter is only 5 kpc in
Abell 262 compared to diameters of $20 - 25$ kpc for the others.  Since
we are assuming that the power output from the radio source is an average,
over many outbursts of the AGN, it may be that a previous outburst of
the radio source in Abell 262 was much more powerful, so that, on average,
the cooling could still be balanced by the input of energy from the
radio source.

\section{Conclusions}
\label{blanton:conclusions}

Central radio-emitting AGN strongly affect the X-ray-emitting gas
in cooling flows.  The radio sources create cavities or ``bubbles'' in the
X-ray gas, which, in turn, confines the radio sources.
In all clusters observed so far, there is no evidence that the radio 
sources are strongly shocking the ICM.  The X-ray-bright shells surrounding
the bubbles are cool, not hot.  Weak shocks may have occurred in the past,
creating the dense shells.  The only evidence for strong-shock heating
of a similar nature has been seen in radio-ISM interactions in galaxies,
and there are only very few cases of this, so far.

The X-ray pressures derived from the shells surrounding the bubbles are
approximately ten times higher than the radio equipartition pressures.
There may be problems with some of the equipartition assumptions, or 
additional contributions to the pressure within the radio bubbles.  One
possibility for this additional pressure source is very hot, diffuse, thermal
gas.

The bubble interiors are less dense than their surroundings and therefore
will rise buoyantly outward into the clusters.  Ghost cavities provide
evidence that this process has occurred.  The buoyantly-rising bubbles
transport energy and magnetic fields into the clusters and they can also
entrain cool gas from the cluster centers outward.

The shell pressures and bubble volumes measured in the X-ray can
be used to determine the total energy output of the radio sources.  A 
radio source repetition rate of $\approx 10^8$ yr is derived from the buoyancy
rise time of the ghost cavities.
A rough comparison of the average energy output of radio sources and the
luminosity of cooling gas shows that the radio sources can supply enough
energy to offset the cooling in cooling flows, at least in some cases.


\acknowledgements
I am very grateful to my collaborators, Craig Sarazin, Brian McNamara, Noam
Soker, Tracy Clarke, and Mike Wise.
Support for E. L. B. was provided by NASA through the {\it Chandra} Fellowship
Program, grant award number PF1-20017, under NASA contract number NAS8-39073.





\end{document}